\documentclass[11pt,a4paper]{article}
\usepackage{amsmath,amssymb}
\usepackage{epsfig,graphicx}
\usepackage{amssymb,graphicx,graphics,epsfig}
\usepackage{latexsym,amsfonts}

\begin{document}

\title{\bf Searches for $W'$ and $Z'$ in models with\\ large extra dimensions}
\author{E.E.~Boos, M.A.~Perfilov, M.N.~Smolyakov, I.P.~Volobuev
\\
\small{\em Skobeltsyn Institute of Nuclear Physics,
Moscow State University} \\
\small{\em 119991 Moscow, Russia} }
\date{}
\maketitle

\begin{abstract}
Characteristic features of processes mediated by gauge bosons are
discussed in the framework of theories with large extra
dimensions. It is shown that if gauge bosons propagate in the
bulk, then there arises a destructive interference not only
between $W$ and $W'$ (or $Z$ and $Z'$), but also between $W'$ and
$Z'$ and the Kaluza-Klein towers of higher excitations of $W$ and
$Z$ bosons respectively. Specific calculations are made and
plotted for the LHC with the center of mass energy 14 TeV.
\end{abstract}

\section{Introduction}
During the last years brane world models with "universal extra
dimensions" are widely discussed in the literature
\cite{Antoniadis:1990ew}-\cite{Burdman:2008qh}. In this case all
the Standard Model fields except the Higgs field can propagate in
the whole multidimensional space-time. This leads to some
interesting phenomenological predictions, which will be discussed
in this paper.

The characteristic feature of theories with compact extra
dimensions is the presence of towers of Kaluza-Klein excitations
of the bulk fields, all the excitations of a bulk field having the
same type of coupling to the fields of the Standard Model. Let us
suppose that, for a bosonic bulk field $\phi$ (or a set of fields)
of arbitrary tensor type in $(4+d)$-dimensional space-time, the
relevant part of the action looks like
\begin{equation}\label{actionKK}
S = \int\sqrt{-\gamma}  d^{4+d}x \, L(\phi) + \int_{\mbox{brane}}
 d^{4}x \,(L_{(SM -\, \phi)}  + g M^{-\frac{d}{2}}J_{SM}*\phi ),
\end{equation}
where $\gamma_{MN} \,(M,N = 0,1,2,3,..., 3+d \, ,  sign \,\gamma =
+,-,\cdots,-)$ denotes the background metric in the bulk,
$L(\phi)$ is the bulk Lagrangian of the field $\phi$, the
Lagrangian of the Standard Model fields, which do not propagate in
the bulk, is denoted by $L_{(SM-\, \phi)}$, the interaction term
$J_{SM}*\phi$ is the scalar product of  the corresponding current
of the Standard Model fields $J_{SM}$ and the field $\phi$ on the
brane, $g$ being a four-dimensional (in general, dimensional)
coupling constant and $M$ being the fundamental energy scale of
the $(4+d)$-dimensional theory defined by the gravitational
interaction; we assume it to be in the TeV energy range.

It is a common knowledge that the bulk field $\phi(x,y), \,
x=\{x^\mu\},y=\{x^i\},  (i=4,..., {3+d})$,  can be expanded in
Kaluza-Klein modes with definite masses $\phi^{(n)}(x)$ and their
wave functions in the space of extra dimension $\psi^{(n)}(y)$ as
follows:
\begin{equation}\label{expandKK}\phi(x,y)= \sum_n \psi^{(n)}(y)
\phi^{(n)}(x), \quad n = (n_1, ..., n_d).
\end{equation}
The current $J_{SM}$ and the coupling constant $g$ are completely
defined by the interaction of  the zero mode $\phi^{(0)}(x)$,
which is a field of the Standard Model or the graviton field, with
the fields of the four-dimensional Standard Model according to
\begin{equation}\label{current}
g J_{SM} = \frac{\delta L^{int}_{SM}}{\delta\phi^{(0)}}\,.
\end{equation}
It is not difficult to show (see a detailed derivation in
\cite{BBSV}) that if we consider this theory for the energy or
momentum transfer much smaller, than the masses of the
Kaluza-Klein excitations $\phi^{(n)},\, n\neq 0$, we can pass to
the effective "low-energy" theory, which can be obtained by the
standard procedure. Namely, we have to drop the momentum
dependence in the propagators of the heavy modes and to integrate
them out in the functional integral built with the original
action. The action of the resulting theory looks like
\begin{eqnarray}\nonumber
S &=&\int  d^{4}x \,\left(\frac{1}{2}\partial_\mu \phi^{(0)}*
\partial^\mu\phi^{(0)} - \frac{1}{2} M_0^2\, \phi^{(0)}* \phi^{(0)}+
L_{int}(\phi^{(0)}) +g M^{-\frac{d}{2}}
\psi^{(0)}(y_b)J_{SM}*\phi^{(0)}+ \right.
\\\label{actionEF} &+& \left.L_{(SM-\, \phi)} + \frac{1}{2} g^2 M^{-{d}}
\left(\sum_{n\neq
0}\frac{(\psi^{(n)}(y_b))^2}{M_n^2}\right)J_{SM}*\Delta*J_{SM}
\right),
\end{eqnarray}
where $M_n$ is the mass of the $n$-th mode and $\Delta$ is the
tensor structure (the numerator) of the propagator with the
momentum equal to zero, which is the same for all modes, $\{y_b\}$
denotes the coordinates of the brane in the space of extra
dimensions. Thus, we get a contact interaction of the Standard
Model fields
\begin{equation}\label{contint}
\lambda J_{SM}*\Delta*J_{SM}, \quad \lambda =\frac{1}{2} g^2
M^{-{d}}\left(\sum_{n \neq 0}\frac{ (\psi^{(n)}(y_b))^2}{M_n^2}
\right),
\end{equation}
the sum of all the other terms in (\ref{actionEF}) being the
Lagrangian of  the Standard Model $L_{SM}$. We see that the
Lagrangian structure is fixed by the corresponding structure of
the Standard Model currents $J_{SM}$ and the spin-density matrix
of the propagating field $\Delta$ defined by the type of the field
$\phi$ as shown in formula (\ref{actionEF}).

\section{Effective Lagrangian for the gauge interaction}
Here we discuss the case of the contact interactions due to the
$SU(2) \times U(1)$ gauge fields in the bulk. These fields are
described in the bulk by vector potentials $W_M$ and $B_M$, which
give rise to four-dimensional vector and scalar fields. The latter
are in the trivial and in the adjoint representations of $SU(2)$
and cannot break $SU(2) \times U(1)$ to $ U(1)_{em}$, as it is
necessary in the Standard Model. For this reason, we assume that
the gauge symmetry is broken in the standard way by the Higgs
field on the brane. It is useful to introduce the charged vector
fields
\begin{equation}
W^{\pm}_\mu = \frac{W^1_\mu \mp W^2_\mu}{\sqrt{2}}
\end{equation}
and the standard mixing of the neutral vector  fields
\begin{eqnarray}\label{mixing}
Z_{\mu} &=& W^3_{\mu} \cos\theta_W - B_{\mu} \sin\theta_W, \\ \nonumber
A_{\mu} &=&  W^3_{\mu} \sin\theta_W + B_{\mu} \cos\theta_W.
\end{eqnarray}
After the spontaneous symmetry breaking  the  neutral component of the
brane Higgs field acquires a vacuum value $v/\sqrt{2}$, and there arises a
quadratic  interaction of the vector fields of the form:
\begin{equation}\label{mix_intW}
\frac{g^2 v^2}{4}M^{-d}\sum_{m,n}\psi_m(y_b)\psi_n(y_b)\,
\eta^{\mu\nu}W^{(m)+}_{\mu} W^{(n)-}_{\nu},
\end{equation}
\begin{equation}\label{mix_intZ}
\frac{(g^2+g'^{2}) v^2}{8}M^{-d}
\sum_{m,n}\psi_m(y_b)\psi_n(y_b)\, \eta^{\mu\nu} Z^{(m)}_{\mu}
Z^{(n)}_{\nu}\,,
\end{equation}
$\psi_m(y_b)$ denoting the wave functions of the Kaluza-Klein
modes of the fields $W_\mu^{\pm}$ and $Z_\mu$ on the brane. Due to
this interaction the Kaluza-Klein modes are no longer the mass
eigenstates; the latter are now superpositions of the modes
\cite{Muck:2002af}. But if the mass scale $gv$ generated by the
Higgs field is much smaller, than the mass of the first
Kaluza-Klein excitation -- and it is exactly the scenario we are
studying -- this mixing of Kaluza-Klein modes can be neglected
\cite{Muck:2002af}. The coupling of the Kaluza-Klein modes to the
fields of the Standard Model is defined by that of the zero mode
and looks like:
\begin{eqnarray}\label{lagr_intV}
L_{int}&=& \frac{g}{\sqrt{2}}
M^{-\frac{d}{2}}\sum_{n>0}\psi_n(y_b)(J^{+\mu }W^{(n)-}_{\mu} +
J^{-\mu }W^{(n)+}_{\mu} ) + \\ \nonumber &+&
\frac{g}{\cos\theta_W}M^{-\frac{d}{2}}\sum_{n>0}\psi_n(y_b)J_{(0)}^\mu
Z^{(n)}_{\mu} + eM^{-\frac{d}{2}}\sum_{n>0}\psi_n(y_b)J_{em}^\mu
A^{(n)}_{\mu},
\end{eqnarray}
where $J^{\pm}_{\mu }$ and $J_{(0)}^\mu$ are the weak charged and
neutral currents of the Standard Model particles and $J_{em}^\mu$
is the electromagnetic current of the Standard Model particles.
Integrating out the heavy modes, we again arrive at the effective
Lagrangian of form (\ref{contint}). Then taking into account that
all the masses are proportional to $M$ and the wave functions are
proportional to $M^{d/2}$, we get the  effective Lagrangian for
the interaction of the Standard Model fields due to the
excitations of the $SU(2)\times U(1)$ gauge bosons
\begin{equation}\label{effl_V}
L_{eff}= \frac{G_F M^2_W }{M^2}\left( C_W J^{+\mu
}J^{-}_{\mu } + C_W J^{-\mu
}J^{+}_{\mu } + C_Z J^{(0)\mu }J^{(0)}_{\mu } +
C_A J^{\mu }_{em}J_{em\,\mu }\right),
\end{equation}
$G_F$ denoting the Fermi constant.
The constants $C_W, C_Z, C_A$ are again model dependent and can be estimated
only in a specific model. In particular, in the simplest model with
two branes and one flat extra dimension the constants can be estimated as
$$
C_W = \frac{\pi^2}{6\sqrt{2}}\, , \quad C_Z = \frac{\sqrt{2}\pi^2}{6
\cos^2\theta_W}\,, \quad C_A = \frac{2\sqrt{2}\pi^2\sin^2\theta_W}{3}\,.
$$

Now let us estimate the constants entering the effective
Lagrangian for the gauge interaction in the case of the the
Randall-Sundrum bulk \cite{Randall:1999ee}. First of all, since
the bulk is 5-dimensional, we can pass to the axial gauge, where
the components corresponding to the extra dimension are equal to
zero \cite{Davoudiasl:1999tf}. Thus, there is no corresponding
scalar fields in the effective four-dimensional theory. The wave
functions $w_n(y)$ of the fields $A^n_\mu(x)$ with definite masses
are solutions of a Sturm-Liouville eigenvalue problem with Neumann
boundary conditions. Due to this fact the wave function of the
massless zero mode, unlike the one for the tensor zero mode, is
constant in the extra dimension. The latter guarantees the
universality of its coupling constant \cite{Rubakov:2001kp}. The
wave functions of the excitations on the brane behave like
$w_n(y)|_{y=y_b} \sim \sqrt{k}$, i.e. similar to the wave
functions of the tensor modes. The masses of the modes appear to
be also in the TeV energy range \cite{Davoudiasl:1999tf}. We will
be interested in the cases where the masses of the modes and the
mass gaps between the modes are quite large, say, of the order of
a few TeV.

Below we will consider some processes with the Kaluza-Klein
electroweak gauge bosons at the energies accessible at the LHC. It
should be noted that the coupling constants and the masses of the
modes depend significantly on the particular model under
consideration. We will also extract the first Kaluza-Klein mode
from the effective Lagrangian (\ref{effl_V}) and suppose that the
accessible energy is above the production threshold of the first
Kaluza-Klein mode. These modes are called $W'$ and $Z'$
respectively. All the other modes will be taken into account by
means of the contact effective interactions.

Symbolic and numerical computations, including simulations of the
Standard Model background for the LHC, have been performed by
means of the CompHEP package \cite{comphep}. The corresponding
Feynman rules have been implemented into the new version of the
CompHEP.

\section{Processes with Kaluza-Klein gauge bosons}
In paper \cite{Boos:2006xe} it was shown in a model independent
way that there exists a nontrivial destructive interference
between the processes mediated by $W$ and $W'$. If we assume that
the gauge bosons propagate in the bulk, then the $W$ boson is just
the zero Kaluza-Klein mode, the $W'$ boson is the first excitation
and there exists an infinite tower of Kaluza-Klein modes above it.
The same is of course valid for $Z$ and $\gamma$. In this case we
expect that the higher Kaluza-Klein modes can also interfere with
the zero and the first modes.

Now let us turn to specific examples. As it was noted in the
previous section, the coupling constants and the masses of the
modes depend significantly on the particular model. For simplicity
we suppose that all the Kaluza-Klein modes have the same coupling
constant as those of the Standard Model $W$, $Z$ bosons and photon
respectively. The masses of the $W'$, $Z'$ bosons and of the first
Kaluza-Klein excitation of the photon are $M_{W'}$, $M_{Z'}$,
$M_{\gamma'}$ respectively. The remaining towers of the modes were
simulated in CompHEP with the help of auxiliary particles with the
masses $M_{W'\_sum}$, $M_{Z'\_sum}$, $M_{\gamma'\_sum}$ and
neglected momentum in the propagators. Indeed, schematically we
can write the amplitude squared as
\begin{eqnarray}\label{amp}
\left|\frac{1}{p^2-M^2}+\frac{1}{p^2-{M'}^2}-\sum_{n=2}^{\infty}\frac{1}{M_{n}^2}\right|^2=\\
\nonumber
\left|\frac{1}{p^2-M^2}+\frac{1}{p^2-{M'}^2}-\frac{1}{M_{sum}^2}\right|^2,
\end{eqnarray}
where $M_{n}$ correspond to the masses of the Kaluza-Klein modes.
The latter formulas show the origin of the parameters which will
be used below. The term $\frac{1}{M_{sum}^2}$ simply corresponds
to the effective contact interaction (\ref{effl_V}).

Now let us consider specific processes including Kaluza-Klein
gauge bosons. For illustrative purposes, all the calculations were
made for the LHC with the center of mass energy
$14\,\textrm{TeV}$.

\begin{figure}[ht]
\begin{center}
\includegraphics[width=11cm]{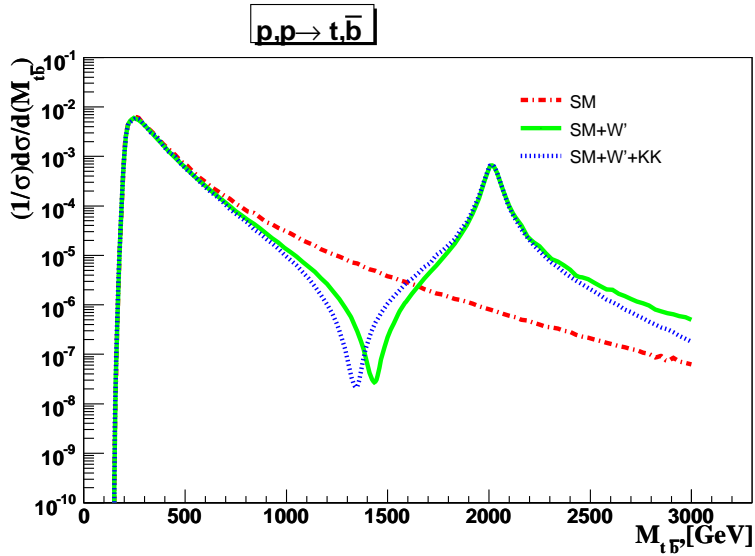}
\caption{\label{bpsv1}Invariant mass distribution for the single
top production at the LHC}
\end{center}
\end{figure}

\begin{figure}[ht]
\begin{center}
\includegraphics[width=11cm]{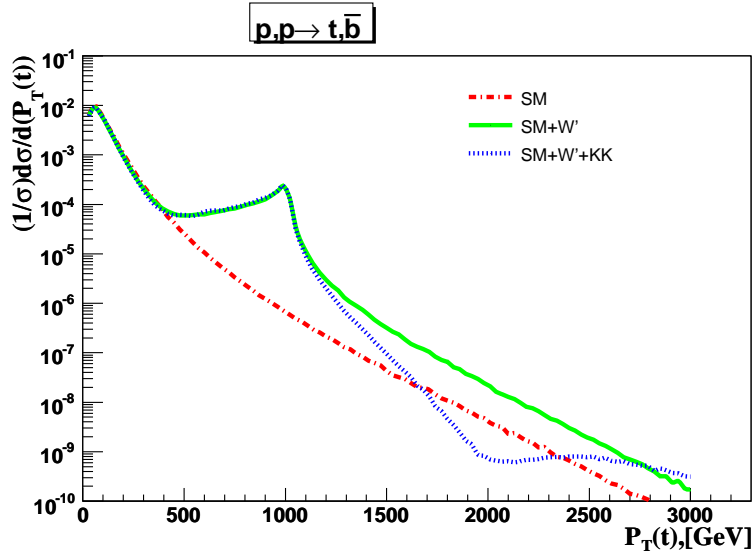}
\caption{\label{bpsv2} $P_{T}$ distribution for the single top
production at the LHC}
\end{center}
\end{figure}

First we consider a process with $W'$ boson plus the remaining
tower of the modes, namely, the single top production. We suppose
that the mass of the first mode $M_{W'}=2\,\textrm{TeV}$, the
effective mass $M_{W'\_sum}=2.8\,\textrm{TeV}$. The width of the
$W'$ resonance has been calculated to be
$\Gamma_{W'}=65.7\,\textrm{GeV}$. The distributions for the
process $u\bar d\to t\bar b$ give the main contribution to the
process $pp\to t\bar b$ at the LHC presented in Figures
\ref{bpsv1} and \ref{bpsv2}.

We made calculations for the Standard Model $W$ boson only, for
the Standard Model $W$ boson plus the $W'$ boson only and for the
Standard Model $W$ boson plus the $W'$ boson plus the remaining
tower of Kaluza-Klein modes. It is clear from Figure \ref{bpsv1}
that the presence of the $W'$ boson leads to a destructive
interference at the energies smaller than the mass of the $W'$
resonance. At the energies larger than the mass of the $W'$
resonance we see that there is an increase of the distributions
tails due to the existence of $W'$ and the corresponding
Kaluza-Klein modes in comparison with the case of the Standard
Model $W$ only (see Figure \ref{bpsv1}). Note that Kaluza-Klein
modes above $W'$ can lead to a quite considerable modification of
the distributions, look at Figure \ref{bpsv2}.

\begin{figure}[ht]
\begin{center}
\includegraphics[width=11cm]{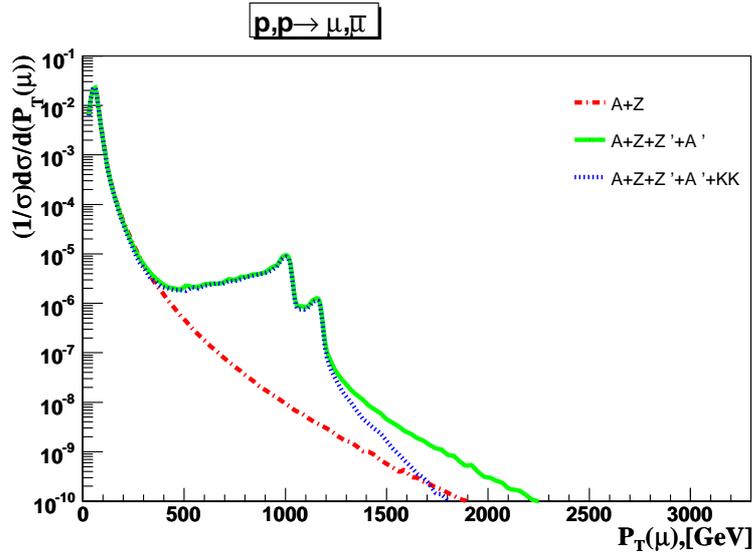}
\caption{\label{bpsv4} $P_{T}$ distribution for the Drell-Yan
process at the LHC}
\end{center}
\end{figure}

Second we consider a process with the $Z'$ boson and the $\gamma'$
boson plus the remaining towers of the modes, namely, the
Drell-Yan process with $u$ quarks, which is also dominant at the
LHC. We suppose that the masses of the first modes are
$M_{Z'}=2.3\,\textrm{TeV}$, $M_{\gamma'}=2\,\textrm{TeV}$, and the
effective masses are $M_{Z'\_sum}=3.2\,\textrm{TeV}$,
$M_{\gamma'\_sum}=2.8\,\textrm{TeV}$.  The widths of the $Z'$ and
$\gamma'$ resonances have been found to be
$\Gamma_{Z'}=0.026\,\textrm{TeV}$ and
$\Gamma_{\gamma'}=0.021\,\textrm{TeV}$ respectively.
\begin{figure}[ht]
\begin{center}
\includegraphics[width=11cm]{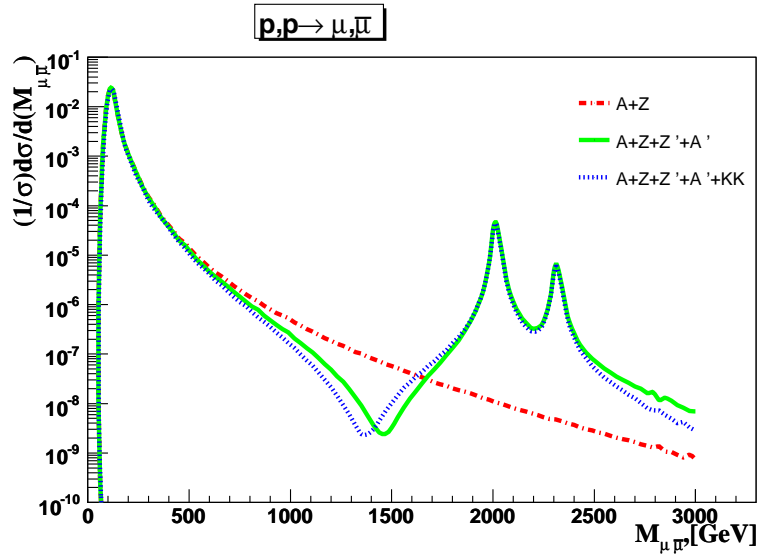}
\caption{\label{bpsv3}Invariant mass distribution for the
Drell-Yan process at the LHC}
\end{center}
\end{figure}
The corresponding distributions are presented in Figures
\ref{bpsv3} and \ref{bpsv4}. One can see analogous properties of
the distributions as those in the case of single top production.

There is a good reason to believe that the NLO corrections do not
destroy this interference picture. First of all, it is clear that
the corrections to the external lines do not alter the structure
of the amplitude  (\ref{amp}). Of course, the most dangerous terms
seem to be  those with the self-energy of the gauge bosons. But
these self-energy terms are defined so as to vanish on the mass
shell and contribute only to the particle widths and to the mass
renormalization.

Thus, our analysis shows that the Kaluza-Klein modes should be
taken into account because they can make contribution to the
amplitudes of the corresponding processes. Of course, in principle
single particles $W''$ or $Z''$ can also provide analogous
effects, but simultaneous effects with Kaluza-Klein gravitons and
$W'$, $Z'$ Kaluza-Klein modes can be interpreted in favor of the
existence of extra dimensions.

{ \large \bf Acknowledgments}

The authors are grateful to V.E.~Bunichev for discussions. The
work was supported by grant of Russian Ministry of Education and
Science NS-4142.2010.2, RFBR grants 08-02-92499-CNRSL$\_$a,
08-02-91002-CERN$\_$a, state contract 02.740.11.0244 and grant
MK-3977.2011.2 of the President of Russian Federation.

\end{document}